# Benchmarking Data Warehouses

Jérôme Darmont, Fadila Bentayeb and Omar Boussaïd

ERIC, University of Lyon 2
5 avenue Pierre Mendès-France
69676 Bron Cedex
France
{jdarmont | boussaid | bentayeb}@eric.univ-lyon2.fr

**Abstract.** Data warehouse architectural choices and optimization techniques are critical to decision support query performance. To facilitate these choices, the performance of the designed data warehouse must be assessed, usually with benchmarks. These tools can either help system users comparing the performances of different systems, or help system engineers testing the effect of various design choices. While the Transaction Processing Performance Council's standard benchmarks address the first point, they are not tunable enough to address the second one and fail to model different data warehouse schemas. By contrast, our Data Warehouse Engineering Benchmark (DWEB) allows generating various ad-hoc synthetic data warehouses and workloads. DWEB is implemented as a Java free software that can be interfaced with most existing relational database management systems. The full specifications of DWEB, as well as experiments we performed to illustrate how our benchmark may be used, are provided in this paper.

## 1. Introduction

When designing a data warehouse, choosing its architecture is crucial. Since it is very dependant on the domain of application and the analysis objectives that are selected for decision support, different solutions are possible. In the ROLAP (Relational On-Line Analytical Process) environment we consider (Chen *et al.*, 2006), the most popular solutions are by far star, snowflake, and constellation schemas (Inmon, 2002; Kimball & Ross, 2002), and other modelling possibilities do exist. This choice of architecture is not neutral: it always has advantages and drawbacks and greatly influences the response time of decision support queries. For example, a snowflake schema with hierarchical dimensions might improve analysis power (namely, when most of the queries are done on the highest hierarchy levels), but induces many more costly join operations than a star schema without hierarchies. Once the architecture is selected, various optimization techniques such as indexing or materializing views further influence querying and refreshing performance. This is espe-cially critical when performing complex tasks, such as computing cubes (Taniar & Rahayu, 2002; Taniar & Tan, 2002; Tan *et al.*, 2003; Taniar *et al.*, 2004) or performing data mining (Tjioe & Taniar, 2005). Again, it is a matter of trade-off between the improvement brought by a given technique and its overhead in terms of maintenance time and additional disk space; and also between different optimization techniques that may cohabit.

To help users make these critical choices of architecture and optimization techniques, the performance of the designed data warehouse needs to be assessed. However, evaluating data warehousing and decision support technologies is an intricate task. Though pertinent, general advice is available, notably on-line (Pendse, 2003; Greenfield, 2004a), more quantitative elements regarding sheer performance are scarce. Thus, we advocate for the use of adapted benchmarks. A benchmark may be defined as a database model and a workload model (set of operations to execute on the database). Different goals may be achieved by using a benchmark: (1) compare the performances of various systems in a given set of experimental conditions (users); (2) evaluate the impact of architectural choices or optimization techniques on the performances of one given system (system designers).

The Transaction Processing Performance Council (TPC[1]), a non-profit organization, defines standard benchmarks and publishes objective and verifiable performance evaluations to the industry. These benchmarks mainly aim at the first benchmarking goal we identified above. However, these benchmarks only model one fixed type of database and they are not very tunable: the only parameter that defines their database is a scale factor setting its size. Nevertheless, in a development context, it is interesting to test a solution (an indexing strategy, for instance) using various database configurations. Furthermore, though there is an ongoing effort at the TPC to design a data warehouse benchmark, the current TPC decision support benchmarks do not properly model a dimensional, star-like schema. They do not address specific warehousing issues such as the ETL (Extract, Transform, Load) process or OLAP (On-Line Analytical Processing) querying either.

Thus, we have proposed a new data warehouse benchmark named DWEB: the *Data Warehouse Engineering Benchmark* (Darmont *et al.*, 2005a). DWEB helps generating ad-hoc synthetic data warehouses (modelled as star, snowflake, or constellation schemas) and workloads, mainly for engineering needs (second benchmarking objective). Thus, DWEB may be viewed more like a benchmark generator than an actual, single benchmark. It is indeed very important to achieve the different kinds of schemas that are used in data warehouses, and to allow designers to select the precise architecture they need to evaluate. This paper expands our previous work along three main axes. First, we present a complete overview of the existing data warehouse benchmarks in this paper. Second, we provide here the full specifications for DWEB, including all its parameters, our query model and the pseudo-code for database and workload generation. We also better detail

---

[1] http://www.tpc.org

DWEB's implementation. Third, we present a new illustration of how our benchmark can be used by evaluating the performance of several index configurations on three test data warehouses.

The remainder of this paper is organized as follows. First, we provide a comprehensive overview of the state of the art regarding decision support benchmarks in Section 2. Then, we motivate the need for a new data warehouse benchmark in Section 3. We detail DWEB's database and workload in Sections 4 and 5, respectively. We also present our implementation of DWEB in Section 6. We discuss our sample experiments with DWEB in Section 7, and finally conclude this paper and provide future research directions in Section 8.

## 2. Existing decision support benchmarks

To the best of our knowledge, relatively few decision support benchmarks have been designed out of the TPC. Some do exist, but their specification is sometimes not fully published (Demarest, 1995). The most notable is presumably the OLAP APB-1 benchmark, which was issued in 1998 by the OLAP council, a now inactive organization founded by four OLAP vendors. APB-1 has been quite extensively used in the late nineties. Its data warehouse schema is architectured around four dimensions: *Customer*, *Product*, *Channel* and *Time*. Its workload of ten queries is aimed at sale forecasting. APB-1 is quite simple and proved limited, since it is not "differentiated to reflect the hurdles that are specific to different industries and functions" (Thomsen, 1998). Finally, some OLAP datasets are also available on-line[2], but they do not qualify as benchmarks, being only raw databases (chiefly, no workload is provided).

In the remainder of this section, we focus more particularly on the TPC benchmarks. The TPC-D benchmark (Ballinger, 1993; Bhashyam, 1996; TPC, 1998) appeared in the mid-nineties, and forms the base of TPC-H and TPC-R that have replaced it (Poess & Floyd, 2000; TPC, 2003). TPC-H and TPC-R are actually identical, only their usage varies. TPC-H is for ad-hoc querying (queries are not known in advance and optimizations are forbidden), while TPC-R is for reporting (queries are known in advance and optimizations are allowed). TPC-H is currently the only decision support benchmark supported by the TPC. TPC-H and TPC-R exploit the same relational database schema as TPC-D: a classical *product-order-supplier* model (represented as a UML class diagram in Figure 1); and the workload from TPC-D supplemented with five new queries. This workload is constituted of twenty-two SQL-92 parameterized, decision-oriented queries labelled Q1 to Q22; and two refresh functions RF1 and RF2 that essentially insert and delete tuples in the ORDER and LINEITEM tables.

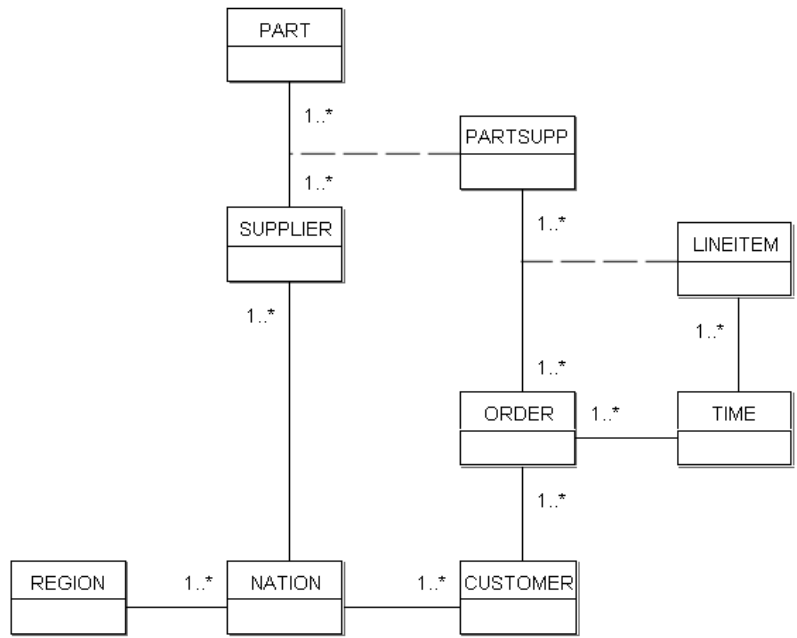

Figure 1: TPC-D, TPC-H, and TPC-R database schema

The query parameters are substituted with the help of a random function following a uniform distribution. Finally, the protocol for running TPC-H or TPC-R includes a load test and a performance test (executed twice), which is further subdivided into a power test and a throughput test. Three primary metrics describe the results in terms of power, throughput, and a composition of the two. Power and throughput are respectively the geometric and arithmetic average of database size divided by execution time.

TPC-DS (Poess *et al.*, 2002), which is currently under development, is the designated successor of TPC-H, and more clearly models a data warehouse. TPC-DS' database schema, whose fact tables are represented in Figure 2, models the decision support functions of a retail product supplier as several snowflake schemas. Catalog and web sales and returns are interrelated, while store management is independent. This model also includes fifteen dimensions that are shared by the fact tables. Thus, the whole model is a constellation schema.

TPC-DS' workload is made of four classes of queries: reporting queries, ad-hoc decision support queries, interactive OLAP queries, and data extraction queries. A set of about five hundred queries is generated from query templates written in SQL-99 – with OLAP extensions (Maniatis *et al.*, 2005). Substitutions on the templates are operated using non-uniform random distributions. The data warehouse maintenance process includes a full ETL process and a specific treatment of the dimensions. For instance, historical dimensions preserve history as new dimension entries are added, while non-historical dimensions do not store aged data any more. Finally, the execution model of TPC-DS consists of four steps: a load test, a query run, a data maintenance run, and another query run. A single throughput metric is proposed, which takes the query and maintenance runs into account.

---

[2] http://cgmlab.cs.dal.ca/downloadarea/datasets/

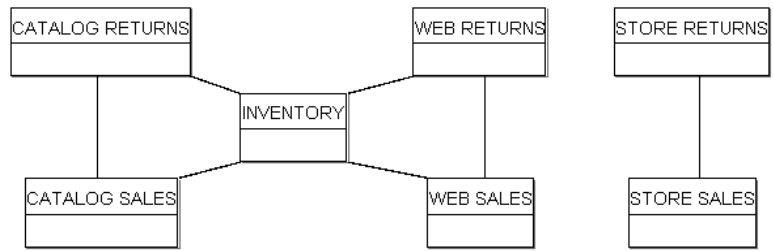

Figure 2: TPC-DS data warehouse schema

## 3. Motivation

Our first motivation to design a data warehouse benchmark is our need to evaluate the efficiency of performance optimization techniques (such as automatic index and materialized view selection techniques) we have been developing for several years. To the best of our knowledge, none of the existing data warehouse benchmarks suits our needs. APB-1's schema is fixed, while we need to test our performance optimization techniques on various data warehouse configurations. Furthermore, it is no longer supported and somewhat difficult to find. TPC-H's database schema, which is inherited from the older and obsolete benchmark TPC-D, is not a dimensional schema such as the typical star schema and its derivatives. Furthermore, its workload, though decision-oriented, does not include explicit OLAP queries either. This benchmark is implicitly considered obsolete by the TPC that has issued some specifications for its successor: TPC-DS. However, TPC-DS has been under development for three years now and is not completed yet. This might be because of its high complexity, especially at the ETL and workload levels.

Furthermore, although the TPC decision support benchmarks are scalable according to the definition of Gray (1993), their schema is fixed. For instance, TPC-DS' constellation schema cannot easily be simplified into a simple star schema. It must be used "as is". Different ad-hoc configurations are not possible. Furthermore, there is only one parameter to define the database, the Scale Factor (*SF*), which sets up its size (from 1 to 100,000 GB). The user cannot control the size of the dimensions and the fact tables separately, for instance. Finally, the user has no control on the workload's definition. The number of generated queries directly depends on *SF* in TPC-DS, for example.

Eventually, in a context where data warehouse architectures and decision support workloads depend a lot on the domain of application, it is very important that designers who wish to evaluate the impact of architectural choices or optimization techniques on global performance can choose and/or compare between several configurations. The TPC benchmarks, which aim at standardized results and propose only one configuration of warehouse schema, are not well adapted to this purpose. TPC-DS is indeed able to evaluate the performance of optimization techniques, but it cannot test their impact on various choices of data warehouse architectures. Generating particular data warehouse configurations (e.g., large-volume dimensions) or ad-hoc query workloads is not possible either, whereas it could be an interesting feature for a data warehouse benchmark.

For all these reasons, we decided to design a full data warehouse benchmark that would be able to model various configurations of database and workload, while being simpler to develop than TPC-DS. In this context (variable architecture, variable size), using a real-life benchmark is not an option. Hence, we designed DWEB for generating various synthetic data warehouses and workloads, which could be individually viewed as full benchmarks.

## 4. DWEB database

### 4.1. Schema

Our design objective for DWEB is to be able to model the different kinds of data warehouse architectures that are popular within a ROLAP environment: classical star schemas, snowflake schemas with hierarchical dimensions, and constellation schemas with multiple fact tables and shared dimensions. To achieve this goal, we propose a data warehouse metamodel (represented as a UML class diagram in Figure 3) that can be instantiated into these different schemas.

We view this metamodel as a middle ground between the multidimensional metamodel from the Common Warehouse Metamodel (CWM) (OMG, 2003; Poole *et al.*, 2003) and the eventual benchmark model. Our metamodel may actually be viewed as an instance of the CWM metamodel, which could be qualified as a meta-metamodel in our context. The upper part of Figure 3 describes a data warehouse (or a datamart, if a datamart is viewed as a small, dedicated data warehouse) as constituted of one or several fact tables that are each described by several dimensions. Each dimension may also describe several fact tables (shared dimensions). Each dimension may be constituted of one or several hierarchies made of different levels. There can be only one level if the dimension is not a hierarchy. Both fact tables and dimension hierarchy levels are relational tables, which are modelled in the lower part of Figure 3. Classically, a table or relation is defined in intention by its attributes and in extension by its tuples or rows. At the intersection of a given attribute and a given tuple lies the value of this attribute in this tuple.

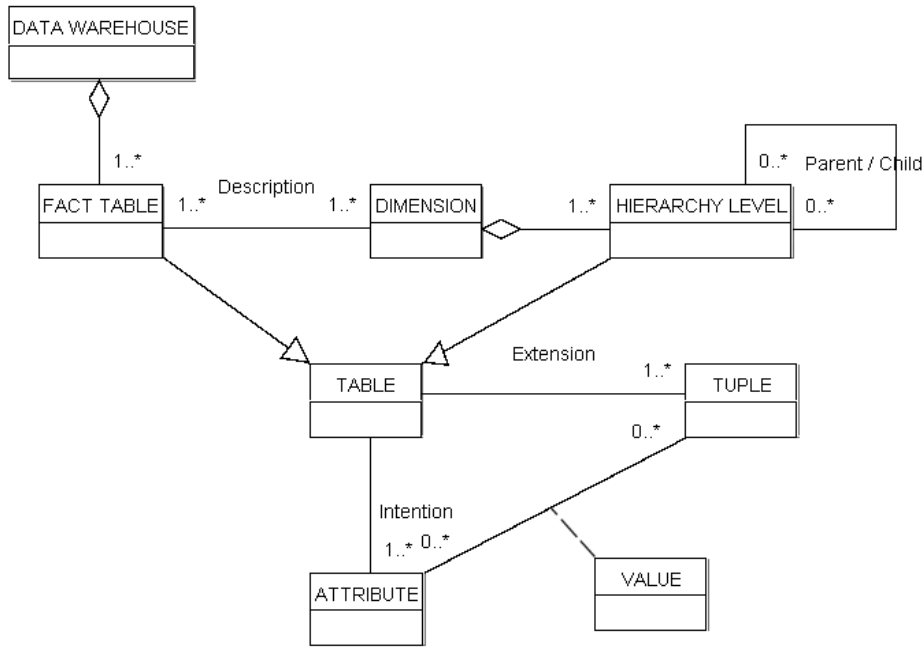

Figure 3: DWEB data warehouse metaschema

Our metamodel is quite simple. It is sufficient to model the data warehouse schemas we aim at (star, snowflake, and constellation schemas), but it is limited and cannot model some particularities that are found in real-life warehouses, such as many-to-many relationships between facts and dimensions, or hierarchy levels shared by several hierarchies. This is currently a deliberate choice, but the metamodel might be extended in the future.

**4.2. Parameterization**

DWEB's database parameters help users selecting the data warehouse architecture they need in a given context. They are aimed at parameterizing the instantiation of the metaschema to produce an actual data warehouse schema. When designing them, we try to meet the four key criteria that make a "good" benchmark, as defined by Gray (1993): *relevance,* the benchmark must answer our engineering needs (as expressed in Section 1); *portability,* the benchmark must be easy to implement on different systems; *scalability,* it must be possible to benchmark small and large databases, and to scale up the benchmark; and *simplicity,* the benchmark must be understandable, otherwise it will not be credible nor used.

Relevance and simplicity are clearly two orthogonal goals. Introducing too few parameters reduces the model's expressiveness, while introducing too many parameters makes it difficult to apprehend by potential users. Furthermore, few of these parameters are likely to be used in practice. In parallel, the generation complexity of the instantiated schema must be mastered. To solve this dilemma, we capitalize on the experience of designing the OCB object-oriented database benchmark

(Darmont & Schneider, 2000). OCB is generic and able to model all the other existing object-oriented database benchmarks, but it is controlled by too many parameters, few of which are used in practice. Hence, we propose to divide the parameter set into two subsets.

The first subset of so-called low-level parameters allows an advanced user to control everything about the data warehouse generation (Table 1). However, the number of low-level parameters can increase dramatically when the schema gets larger. For instance, if there are several fact tables, all their characteristics, including dimensions and their own characteristics, must be defined for each fact table.

| Parameter name | Meaning |
| --- | --- |
| NB_FT | Number of fact tables |
| NB_DIM(f) | Number of dimensions describing fact table #f |
| TOT_NB_DIM | Total number of dimensions |
| NB_MEAS(f) | Number of measures in fact table #f |
| DENSITY(f) | Density rate in fact table #f |
| NB_LEVELS(d) | Number of hierarchy levels in dimension #d |
| NB_ATT(d,h) | Number of attributes in hierarchy level #h of dimension #d |
| HHLEVEL_SIZE(d) | Cardinality of the highest hierarchy level of dimension #d |
| DIM_SFACTOR(d) | Size scale factor in the hierarchy levels of dimension #d |

Table 1: DWEB warehouse low-level parameters

Thus, we designed a layer above with much fewer parameters that may be easily understood and set up (Table 2). More precisely, these high-level parameters are average values for the low-level parameters. At database generation time, the high-level parameters are exploited by random functions (following a Gaussian distribution) to automatically set up the low-level parameters. Finally, unlike the number of low-level parameters, the number of high-level parameters always remains constant and reasonable (less than ten parameters).

| Parameter name | Meaning | Default value |
| --- | --- | --- |
| AVG_NB_FT | Average number of fact tables | 1 |
| AVG_NB_DIM | Average number of dimensions per fact table | 5 |
| AVG_TOT_NB_DIM | Average total number of dimensions | 5 |
| AVG_NB_MEAS | Average number of measures in fact tables | 5 |
| AVG_DENSITY | Average density rate in fact tables | 0.6 |
| AVG_NB_LEVELS | Average number of hierarchy levels in dimensions | 3 |
| AVG_NB_ATT | Average number of attributes in hierarchy levels | 5 |
| AVG_HHLEVEL_SIZE | Average cardinality of the highest hierarchy levels | 10 |
| DIM_SFACTOR | Average size scale factor within hierarchy levels | 10 |

Table 2: DWEB warehouse high-level parameters

Users may choose to set up either the full set of low-level parameters, or only the high-level parameters, for which we propose default values that correspond to a snowflake schema. Note that these parameters control both schema and data generation.

*Remarks:*

- Since shared dimensions are possible, $TOT\_NB\_DIM \leq \sum_{i=1}^{NB\_FT} NB\_DIM(i)$.

- The cardinal of a fact table is usually lower or equal to the product of its dimensions' cardinals. This is why we introduce the notion of density. A density rate of one indicates that all the possible combinations of the dimension primary keys are present in the fact table. When the density rate decreases, we progressively eliminate some of these combinations (see Section 4.3).

- This parameter helps controlling the size of the fact table, independently of the size of its dimensions, which are defined by the *HHLEVEL_SIZE* and *DIM_SFACTOR* parameters (see below).

- Within a dimension, a given hierarchy level normally has a greater cardinality than the next level. For example, in a *town-region-country* hierarchy, the number of towns must be greater than the number of regions, which must be in turn greater than the number of countries. Furthermore, there is often a significant scale factor between these cardinalities (e.g., one thousand towns, one hundred regions, ten countries). Hence, we model the cardinality of hierarchy levels by assigning a "starting" cardinality to the highest level in the hierarchy (*HHLEVEL_SIZE*), and then by multiplying it by a predefined scale factor (*DIM_SFACTOR*) for each lower-level hierarchy.

- The global size of the data warehouse is assessed at generation time (see Section 6) so that the user retains full control over it.

**4.3. Generation algorithm**

The instantiation of the DWEB metaschema into an actual benchmark schema is done in two steps: (1) build the dimensions; (2) build the fact tables. The pseudo-code for these two steps is provided in Figures 4 and 5, respectively. Each of these steps is further subdivided, for each dimension or each fact table, into generating its intention and extension. In addition, hierarchies of dimensions must be managed. Note that they are generated starting from the highest level of hierarchy. For instance, for our *town-region-country* sample hierarchy, we build the country level first, then the region level, and eventually the town level. Hence, tuples from a given hierarchy level can refer to tuples from the next level (that are already created) with the help of a foreign key.

```
For i = 1 to TOT_NB_DIM do
    previous_ptr = NIL
    size = HHLEVEL_SIZE(i)
    For j = 1 to NB_LEVELS(i) do
        // Intention
        h1 = New(Hierarchy_level)
        h1.intention = Primary_key()
        For k = 1 to NB_ATT(i,j) do
            h1.intention = h1.intention ∪ String_descriptor()
        End for
        // Hierarchy management
        h1.child = previous_ptr
        h1.parent = NIL
        If previous_ptr ≠ NIL then
            previous_ptr.parent = h1
            h1.intention = h1.intention
                ∪ previous_ptr.intention.primary_key // Foreign key
        End if
        // Extension
        h1.extension = ∅
        For k = 1 to size do
            new_tuple = Integer_primary_key()
            For l = 1 to NB_ATT(i,j) do
                new_tuple = new_tuple ∪ Random_string()
            End for
            If previous_ptr ≠ NIL then
                new_tuple = new_tuple ∪ Random_key(previous_ptr)
            End if
            h1.extension = h1.extension ∪ new_tuple
        End for
        previous_ptr = h1
        size = size * DIM_SFACTOR(i)
    End for
    dim(i) = h1 // First (lowest) level of the hierarchy
End for
```

Figure 4: DWEB dimensions generation algorithm

```
For i = 1 to TOT_NB_FT do
    // Intention
    ft(i).intention = ∅
    For k = 1 to NB_DIM(i) do
        j = Random_dimension(ft(i))
        ft(i).intention = ft(i).intention ∪ ft(i).dim(j).primary key
    End for
    For k to NB_MEAS(i) do
        ft(i).intention = ft(i).intention ∪ Float_measure()
    End for
    // Extension
    ft(i).extension = ∅
    For j = 1 to NB_DIM(i) do // Cartesian product
        ft(i).extension = ft(i).extension × ft(i).dim(j).primary key
    End for
    to_delete = DENSITY(i) * |ft(i).extension)|
    For j = 1 to to_delete do
        Random_delete(ft(i).extension)
    End for
    For j = 1 to |ft(i).extension)| do // With |ft(i).extension)| updated
        For k = 1 to NB_MEAS(i) do
            Ft(i).extension.tuple(j).measure(k) = Random_float()
        End for
    End for
End for
```

Figure 5: DWEB fact tables generation algorithm

We use three main classes of functions and one procedure in these algorithms.

1. `Primary_key()`, `String_descriptor()` and `Float_measure()` return attribute names for primary keys, descriptors in hierarchy levels, and measures in fact tables, respectively. These names are labelled sequentially and prefixed by the table's name (e.g., DIM1_1_DESCR1, DIM1_1_DESCR2...).
2. `Integer_primary_key()`, `Random_key()`, `Random_string()` and `Random_float()` return sequential integers with respect to a given table (no duplicates are allowed), random instances of the specified table's primary key (random values for a foreign key), random strings of fixed size (20 characters) selected from a precomputed referential of strings and prefixed by the corresponding attribute name, and random single-precision real numbers, respectively.
3. `Random_dimension()` returns a dimension that is chosen among the existing dimensions that are not already describing the fact table in parameter.
4. `Random_delete()` deletes one tuple at random from the extension of a table.

Except in the `Random_delete()` procedure, where the random distribution is uniform, we use Gaussian random distributions to introduce a skew, so that some of the data, whether in the fact tables or the dimensions, are referenced more frequently than others as it is normally the case in real-life data warehouses.

*Remark:* The way density is managed in Figure 5 is grossly non-optimal. We chose to present the algorithm that way for the sake of clarity, but the actual implementation does not create all the tuples from the Cartesian product, and then delete some of them. It directly generates the right number of tuples by using the density rate as a probability for each tuple to be created.

## 5. DWEB workload

In a data warehouse benchmark, the workload may be subdivided into: (1) a load of decision support queries (mostly OLAP queries); (2) the ETL (data generation and maintenance) process. To design DWEB's workload, we inspire both from TPC-DS' workload definition (which is very elaborate) and information regarding data warehouse performance from other sources (BMC, 2000; Greenfield, 2004b). However, TPC-DS' workload is quite complex and somehow confusing. The reporting, ad-hoc decision support and OLAP query classes are very similar, for instance, but none of them include any specific OLAP operator such as Cube or Rollup (Tan *et al.*, 2004). Since we want to meet Gray's simplicity criterion, we propose a simpler workload. In particular, we do not address the issue of nested queries for now. Furthermore, we also have to design a workload that is consistent with the variable nature of the DWEB data warehouses.

We also, in a first step, mainly focus on the definition of a query model that excludes update operations. Modelling the full ETL and warehouse refreshing processes is a complex task requiring processes that are out of the scope of this work (Schlesinger *et al.*, 2005). Hence, we postpone this for now. We consider that the current DWEB specifications provide a raw loading evaluation framework. The DWEB database may indeed be generated into flat files, and then loaded into a data warehouse using the ETL tools provided by the system.

## 5.1. Query model

The DWEB workload models two different classes of queries: purely decision-oriented queries involving common OLAP operations, such as cube, roll-up, drill down and slice and dice; and extraction queries (simple join queries). We define our generic query model (Figure 6) as a grammar that is a subset of the SQL-99 standard, which introduces much-needed analytical capabilities to relational database querying. This increases the ability to perform dynamic, analytic SQL queries.

```
Query ::-

Select                  ![<Attribute Clause> | <Aggregate Clause>
                        | [<Attribute Clause>, <Aggregate Clause>]]
From                    !<Table Clause> [<Where Clause>
                        || [<Group by Clause> * <Having Clause>]]

Attribute Clause ::-    Attribute name [[, <Attribute Clause>] | ⊥]
Aggregate Clause ::-    ![Aggregate function name (Attribute name)] [As Alias]
                        [[, <Aggregate Clause>] | ⊥]

Table Clause ::-        Table name [[, <Table Clause>] | ⊥]

Where Clause ::-        Where ![<Condition Clause> | <Join Clause>
                        | [<Condition Clause> And <Join Clause>]]
Condition Clause ::-    ![Attribute name <Comparison operator> <Operand Clause>]
                        [[<Logical operator> <Condition Clause>] | ⊥]
Operand Clause ::-      [Attribute name | Attribute value | Attribute value list]
Join Clause ::-         ![Attribute name i = Attribute name j]
                        [[And <Join Clause>] | ⊥]

Group by Clause ::-     Group by [Cube | Rollup] <Attribute Clause>
Having Clause ::-       [Alias | Aggregate function name (Attribute name)]
                        <Comparison operator> [Attribute name | Attribute value list]

Key:                    The [ and ] brackets are delimiters.
                        !<A>: A is required.
                        *<A>: A is optional.
                        <A || B>: A or B.
                        <A | B>: A exclusive or B.
                        ⊥: empty clause.
                        SQL language elements are indicated in bold.
```

Figure 6: DWEB query model

## 5.2. Parameterization

DWEB's workload parameters help users tailoring the benchmark's load, which is also dependent from the warehouse schema, to their needs. Just like DWEB's database parameter set (Section 4.2), DWEB's workload parameter set (Table 3) has been designed with Gray's simplicity criterion in mind. These parameters determine how the query model from Figure 6 is instantiated. These parameters help defining the workload's size and complexity, by setting up the proportion of complex OLAP queries (i.e., the class of queries) in the workload , the number of aggregation operations, the presence of a Having clause in the query, or the number of subsequent drill down operations.

Here, we have only a limited number of high-level parameters (eight parameters, since *PROB_EXTRACT* and *PROB_ROLLUP* are derived from *PROB_OLAP* and *PROB_CUBE*, respectively). Indeed, it cannot be envisaged to dive further into detail if the workload is as large as several hundred queries, which is quite typical.

| Parameter name | Meaning | Default value |
| --- | --- | --- |
| NB_Q | Approximate number of queries in the workload | 100 |
| AVG_NB_ATT | Average number of selected attributes in a query | 5 |
| AVG_NB_RESTR | Average number of restrictions in a query | 3 |
| PROB_OLAP | Probability that the query type is OLAP | 0.9 |
| PROB_EXTRACT | Probability that the query is an extraction query | 1 - PROB_OLAP |
| AVG_NB_AGGREG | Average number of aggregations in an OLAP query | 3 |
| PROB_CUBE | Probability of an OLAP query to use the Cube operator | 0.3 |
| PROB_ROLLUP | Probability of an OLAP query to use the Rollup operator | 1 - PROB_CUBE |
| PROB_HAVING | Probability of an OLAP query to include an Having clause | 0.2 |
| AVG_NB_DD | Average number of drill downs after an OLAP query | 3 |

Table 3: DWEB workload parameters

*Remark: NB_Q* is only an *approximate* number of queries because the number of drill down operations after an OLAP query may vary. Hence we can stop generating queries only when we actually have generated as many or more queries than *NB_Q*.

## 5.3. Generation algorithm

The pseudo-code of DWEB's workload generation algorithm is presented in Figure 7. The algorithm's purpose is to generate a set of SQL-99 queries that can be directly executed on the synthetic data warehouse defined in Section 4. It is subdivided into two steps: (1) generate an initial query that may either be an OLAP or an extraction (join) query; (2) if the initial query is an OLAP query, execute a certain number of drill down operations based on the first OLAP query. More precisely, each time a drill down is performed, an attribute from a lower level of dimension hierarchy is added to the attribute clause of the previous query.

Step 1 is further subdivided into three substeps: (1) the Select, From, and Where clauses of a query are generated simultaneously by randomly selecting a fact table and dimensions, including a hierarchy level within a given dimension hierarchy; (2) the Where clause is supplemented with additional conditions; (3) eventually, it is decided whether the query is an OLAP query or an extraction query. In the second case, the query is complete. In the first case, aggregate functions applied to measures of the fact table are added in the query, as well as a Group by clause that may include either the Cube or the Rollup operator. A Having clause may optionally be added in too. The aggregate function we apply on measures is always Sum since it is the most common aggregate in cubes. Furthermore, other aggregate functions bear similar time complexities, so they would not bring in any more insight in a performance study.

```
n = 0
While n < NB_Q do
    // Step 1: Initial query
    // Step 1.2: Select, From and Where clauses
    i = Random_FT() // Fact table selection
    attribute_list = Ø
    table_list = ft(i)
    condition_list = Ø
    For k = 1 to Random_int(AVG_NB_ATT) do
        j = Random_dimension(ft(i)) // Dimension selection
        l = Random_int(1, ft(i).dim(j).nb_levels)
        // Positioning on hierarchy level l
        hl = ft(i).dim(j) // Current hierarchy level
        m = 1 // Level counter
        fk = ft(i).intention.primary_key.element(j)
        // This foreign key corresponds to ft(i).dim(j).primary_key
        While m < l and hl.child ≠ NIL do
            // Build join
            table_list = table_list ∪ hl
            condition_list = condition_list
                ∪ (fk = hl.intention.primary_key)
            // Next level
            fk = hl.intention.foreign_key
            m = m + 1
            hl = hl.child
        End while
        attribute_list = attribute_list ∪ Random_attribute(hl.intention)
    End for
    // Step 1.2: Supplement Where clause
    For k = 1 to Random_int(AVG_NB_RESTR) do
        condition_list = condition_list
            ∪ (Random_attribute(attribute_list) = Random_string())
    End for
    // Step 1.3: OLAP or extraction query selection
    p1 = Random_float(0, 1)
    If p1 ≤ PROB_OLAP then // OLAP query
        // Aggregate clause
        aggregate_list = Ø
        For k = 1 to Random_int(AVG_NB_AGGREG) do
            aggregate_list = aggregate_list
                ∪ (Random_measure(ft(i).intention)
        End for
        // Group by clause
```

```
            group_by_list = attribute_list
            p2 = Random_float(0, 1)
            If p2 ≤ PROB_CUBE then
                group_by_operator = CUBE
            Else
                group_by_operator = ROLLUP
            End if
            // Having clause
            P3 = Random_float(0, 1)
            If p3 ≤ PROB_HAVING then
                having_clause
                    = (Random_attribute(aggregate_list), ≥, Random_float())
            Else
                having_clause = ∅
            End if
        Else // Extraction query
            group_by_list = ∅
            group_by_operator = ∅
            having_clause = ∅
        End if
        // SQL query generation
        Gen_query(attribute_list, aggregate_list, table_list, condition_list,
            group_by_list, group_by_operator, having_clause)
        n = n + 1
        // Step 2: Possible subsequent DRILL DOWN queries
        If p1 ≤ PROB_OLAP then
            k = 0
            While k < Random_int(AVG_NB_DD) and hl.parent ≠ NIL do
                k = k + 1
                hl = hl.parent
                att = Random_attribute(hl.intention)
                attribute_list = attribute_list ∪ att
                group_by_list = group_by_list ∪ att
                Gen_query(attribute_list, aggregate_list, table_list,
                    condition_list,  group_by_list, group_by_operator,
                    having_clause)
            End while
            n = n + k
        End if
End while
```

Figure 7: DWEB workload generation algorithm

We use three classes of functions and a procedure in this algorithm.

1. `Random_string()` and `Random_float()` are the same functions than those already described in Section 4.3. However, we introduce the possibility for `Random_float()` to use either a uniform or a Gaussian random distribution. This depends on the function parameters: either a range of values (uniform) or an average value (Gaussian). Finally, we introduce the `Random_int()` function that behaves just like `Random_float()` but returns integer values.

2. `Random_FT()` and `Random_dimension()` help selecting a fact table or a dimension describing a given fact table, respectively. They both use a Gaussian random distribution, which introduces an access skew at the fact table and dimension levels. `Random_dimension()` is also already described in Section 4.3.

3. `Random_attribute()` and `Random_measure()` are very close in behaviour. They return an attribute or a measure, respectively, from a table intention or a list of attributes. They both use a Gaussian random distribution.
4. `Gen_query()` is the procedure that actually generates the SQL-99 code of the workload queries, given all the parameters that are needed to instantiate our query model.

## 6. DWEB implementation

DWEB is implemented as a Java software. We selected the Java language to meet Gray's portability requirement. The current version of our prototype is able to generate star, snowflake, and constellation schemas, and suitable workloads for these schemas. Its only limitation with respect to our metamodel is that it cannot generate several distinct hierarchies for the same dimension. Furthermore, since DWEB's parameters might sound abstract, our prototype provides an estimation of the data warehouse size in megabytes after they are set up and before the database is generated. Hence, users can adjust the parameters to better represent the kind of warehouse they need.

The interface of our Java application is actually constituted of two GUIs (Graphical User Interfaces). The first one is the *Generator*, the core of DWEB. It actually implements all the algorithms provided in Sections 4.3 and 5.3 and helps selecting either low or high-level parameters and generating any data warehouse and corresponding workload (Figure 8-a). Data warehouses are currently directly loaded into a database management system (DMBS), but we also plan to save them as files of SQL queries to better evaluate the loading phase. Workloads are already saved as files of SQL queries. The second GUI, the *Workload executor*, helps connecting to an existing data warehouse and running an existing workload on it (Figure 8-b). The execution time for each query is recorded separately and can be exported in a CSV file that can later be processed in a spreadsheet or any other application. Both GUIs can be interfaced with most existing relational database management systems through JDBC.

Our software is constantly evolving. For example, since we use a lot of random functions, we plan to include in our prototype a better than standard pseudorandom number generator, such as the Lewis and Payne (1973) generator, which has a huge period, or the Mersenne Twister (Matsumoto & Nishimura, 1998), which is currently one of the best pseudorandom number generators. However, the latest version of DWEB is always freely available on-line[3].

---
[3] http://bdd.univ-lyon2.fr/download/dweb.tgz

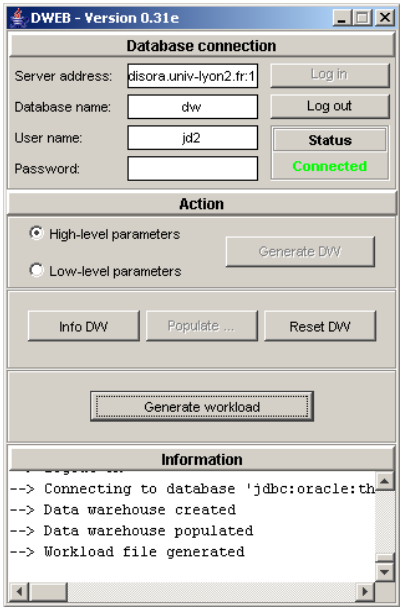 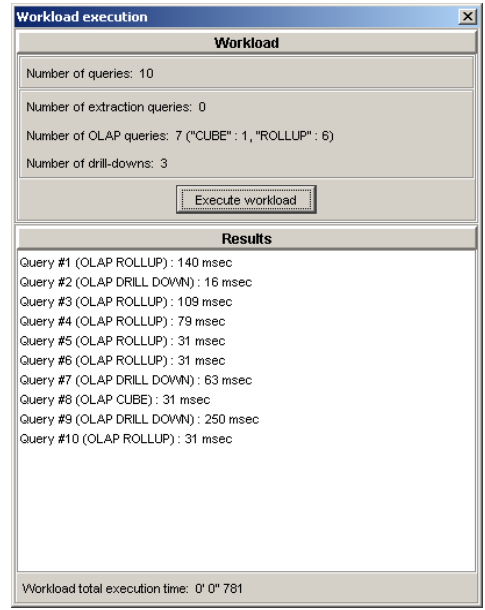

(a) Generator                                    (b) Workload executor

Figure 8: DWEB GUIs

# 7. Sample usage of DWEB

## 7.1. Experiments scope

In order to illustrate one possible usage for DWEB, we evaluate the efficiency of several indexing techniques on several configurations of warehouses. Since there is presumably no perfect index for all the ROLAP logical data warehouse models, we aim at verifying which indices work best on a given schema type. To achieve this goal, we generate with DWEB three test data warehouses labelled DW1 to DW3 and their associated workloads. DW1 and DW2 are modelled as snowflake schemas, and DW3 as a star schema. Then, we successively execute the corresponding workloads on these data warehouses using four index configurations labelled IC0 to IC3. IC0 actually uses no index and serves as a reference.

Index configuration IC1 is constituted of bitmap join indices (O'Neil & Graefe, 1995) built on the fact tables and the dimensions' lowest hierarchy levels (i.e., only on the central star in the snowflake schemas of DW1 and DW2). Bitmap join indices are well suited to the data warehouse environment. They indeed improve the response time of such common operations as And, Or, Not, or Count that can operate on the bitmaps (and thus directly in memory) instead of the source data. Furthermore, joins are computed *a priori* when the indices are created. Index configuration IC2 adds to IC1 bitmap join indices between the dimensions' hierarchy levels. If course, IC2 is not applied to the DW3 data warehouse, which is modelled as a star schema. Finally, index configuration IC3 is made of a star join index (Bellatreche *et al.*, 2002). Such an index may link all the dimensions to the fact table. It is then said whole and may benefit to any query on the star schema. Its storage space is very large, though. A partial star join index may be built on the fact table and only several dimensions. However, we used only whole star join indices in this study to maximize performance improvement.

Note that we do not expect to achieve new results with these experiments. What we seek to do is providing an example of how DWEB may be used, and demonstrating that the results it provides are consistent with the previous results achieved by data warehouse indices' designers (O'Neil & Graefe, 1995; Bellatreche *et al.*, 2002).

**7.2. Hardware and software configuration**

Our tests have been performed on a Centrino 1.7 GHz PC with 1024 MB of RAM running Windows XP and Oracle 10g. All the experiments have been run "locally", i.e., the Oracle server and client were on the same machine, so that network latency did not interfere with the results.

**7.3. Benchmark configuration**

DW1's snowflake schema is constituted of one fact table and two dimensions. The DWEB low-level parameters that define it are displayed in Table 4. Its schema is showed as a UML class diagram in Figure 9. Its actual size is 92.9 MB. DW2's snowflake schema is constituted of one fact table and four dimensions. The DWEB low-level parameters that define it are displayed in Table 5. Its schema is showed as a UML class diagram in Figure 10. Its actual size is 224.5 MB. Finally, DW3's star schema is constituted of one fact table and three dimensions. The DWEB low-level parameters that define it are displayed in Table 6. Its schema is showed as a UML class diagram in Figure 11. Its actual size is 68.3 MB.

For each data warehouse, we generate a workload of twenty queries ($NB\_Q = 20$). The other parameters are set up to the default values specified in Table 3. The queries to be executed on data warehouse DWi are labelled Qi.1 to Qi.20. Due to space constraints, we cannot include these three workloads in this paper, but their SQL code is available on-line[4].

---

[4] http://bdd.univ-lyon2.fr/documents/dweb-workloads.pdf

| Parameter | Value |
|---|---|
| NB_FT | 1 |
| NB_DIM(1) | 2 |
| TOT_NB_DIM | 2 |
| NB_MEAS(1) | 5 |
| DENSITY(1) | 0.6 |
| NB_LEVELS(1) | 2 |
| NB_ATT(1) | 5 / 5 |
| NB_LEVELS(2) | 3 |
| NB_ATT(2) | 4 / 4 / 4 |
| HHLEVEL_SIZE(1-2) | 18 |
| DIM_SFACTOR(1-2) | 18 |

Table 4: DW1 parameters

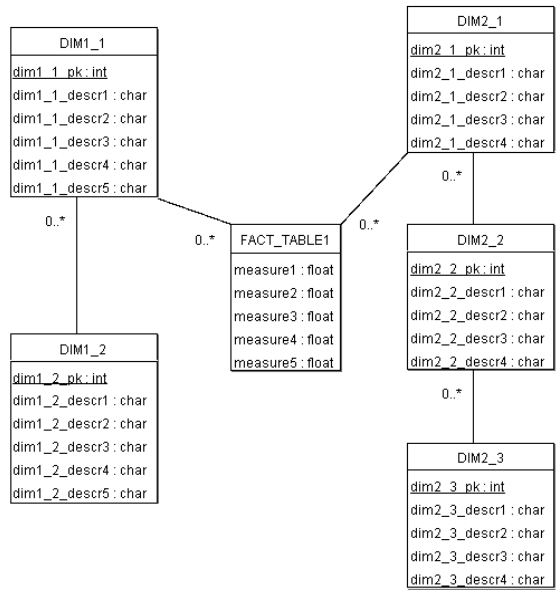

Figure 9: DW1 snowflake schema

| Parameter | Value |
|---|---|
| NB_FT | 1 |
| NB_DIM(1) | 4 |
| TOT_NB_DIM | 4 |
| NB_MEAS(1) | 3 |
| DENSITY(1) | 0.25 |
| NB_LEVELS(1) | 1 |
| NB_ATT(1) | 4 |
| NB_LEVELS(2) | 2 |
| NB_ATT(2) | 2 / 3 |
| NB_LEVELS(3) | 3 |
| NB_ATT(3) | 3 / 3 / 2 |
| NB_LEVELS(4) | 3 |
| NB_ATT(4) | 2 / 2 / 3 |
| HHLEVEL_SIZE(1-4) | 8 |
| DIM_SFACTOR(1-4) | 5 |

Table 5: DW2 parameters

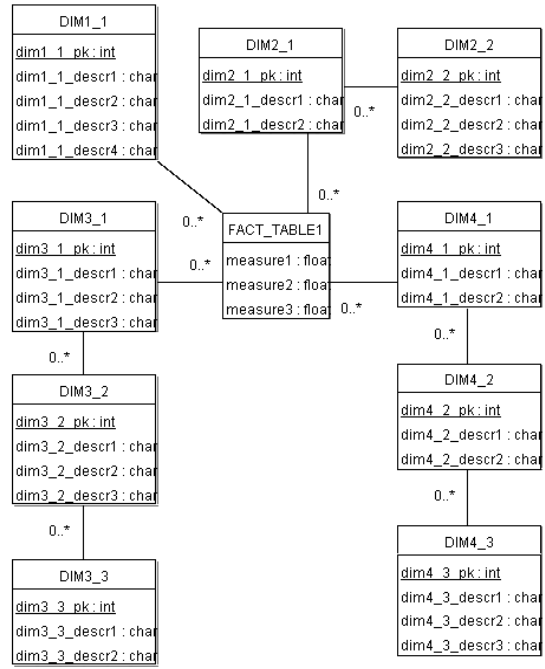

Figure 10: DW2 snowflake schema

| Parameter | Value |
|---|---|
| NB_FT | 1 |
| NB_DIM(1) | 3 |
| TOT_NB_DIM | 3 |
| NB_MEAS(1) | 5 |
| DENSITY(1) | 0.8 |
| NB_LEVELS(1-3) | 1 |
| NB_ATT(1-3) | 5 |
| HHLEVEL_SIZE(1-2) | 100 |
| HHLEVEL_SIZE(3) | 70 |
| DIM_SFACTOR(1-3) | n/a |

Table 6: DW3 parameters

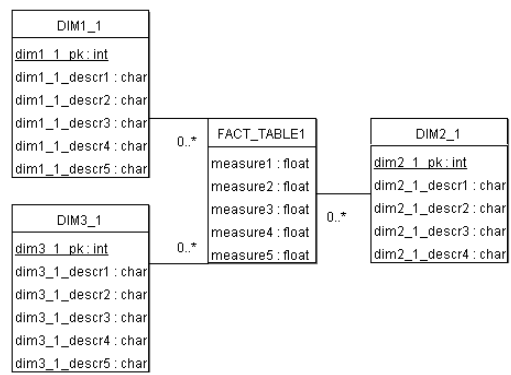

Figure 11: DW3 star schema

## 7.4. Results

Since we are chiefly interested in raw performance in these experiments, execution time is the only metric we selected. However, we do envisage more elaborate metrics (Section 8). Table 7 presents the execution time (in milliseconds) of each of the queries Q1.1 to Q1.20 on data warehouse DW1, using each of the index configurations IC0 to IC3. Table 7's last line also features the average gain in performance when using a given index configuration instead of no index.

| Query | IC0 | IC1 | IC2 | IC3 |
|---|---|---|---|---|
| Q1.1 | 120 574 | 115 926 | 121 074 | 197 774 |
| Q1.2 | 51 133 | 34 981 | 31 105 | 66 716 |
| Q1.3 | 95 618 | 37 954 | 42 861 | 66 275 |
| Q1.4 | 74 958 | 30 564 | 29 222 | 36 393 |
| Q1.5 | 2 556 075 | 1 130 315 | 1 300 580 | 3 181 364 |
| Q1.6 | 38 255 | 74 898 | 50 403 | 101 486 |
| Q1.7 | 391 | 90 | 160 | 601 |
| Q1.8 | 75 999 | 117 179 | 221 889 | 131 359 |
| Q1.9 | 12 228 | 11 486 | 13 720 | 15 162 |
| Q1.10 | 808 402 | 604 980 | 633 371 | 1 263 407 |
| Q1.11 | 4 577 | 4 326 | 6 098 | 4 847 |
| Q1.12 | 105 952 | 27 230 | 42 942 | 46 937 |
| Q1.13 | 1 618 317 | 944 818 | 990 104 | 1 052 303 |
| Q1.14 | 1 461 492 | 1 050 120 | 1 392 512 | 1 022 901 |
| Q1.15 | 59 946 | 81 898 | 66 886 | 207 719 |
| Q1.16 | 324 256 | 343 894 | 242 419 | 494 120 |
| Q1.17 | 835 141 | 705 024 | 677 003 | 2 199 853 |
| Q1.18 | 2 414 913 | 1 731 830 | 2 760 129 | 5 063 301 |
| Q1.19 | 313 560 | 261 286 | 526 998 | 317 437 |
| Q1.20 | 577 462 | 384 673 | 481 551 | 814 208 |
| **Gain** | **0%** | **33.4%** | **16.6%** | **-41.0%** |

Table 7: DW1 results

Table 7 first shows that index configuration IC1 noticeably improves response time, especially for queries that return large results (Q1.5, Q1.13, Q1.14, and Q1.18). Using no index is better only for some shorter queries such as Q1.6 or Q1.15, but in these cases, it is not penalizing since response times are low and the difference in performance is small too. Bitmap join indices are thus experimented to be the most useful when queries return large results. We can also notice on Table 7 that adding bitmap join indices between the dimensions' hierarchy levels (index configuration IC2) degrades the performances. They indeed incur many index scans, whereas the dimensions' highest hierarchy tables have a relatively small size that does not actually justify indexing. Finally, the star join index (IC3) appears completely ill-suited to the snowflake schema of DW1 and clearly degrades the performances, especially for queries that return large results, whose response time may triple. This was expected, since star join indices are aimed at accelerating queries formulated on a star schema only, but maybe not to this extent.

Table 8 presents the execution time (in milliseconds) of each of the queries Q2.1 to Q2.20 on data warehouse DW2, using each of the index configurations IC0 to IC3. Table 8's last line also features the average gain in performance when using a given index configuration instead of no index. Table 8's results basically confirm those from Table 7. However, the effects of indices are significantly softened. DW2's fact table is indeed thrice as large as DW1's, while being much sparser (its density is more than twice lower than DW1's). Hence, bitmap join indices (configurations IC1 and IC2) are at the same time bigger and less pertinent when computing sparse cubes. This low density in the fact table also reduces the bad performances of the star join index (configuration IC3), since links to the dimensions are less numerous.

| Query | IC0 | IC1 | IC2 | IC3 |
|---|---|---|---|---|
| Q2.1 | 14 351 | 14 701 | 13 279 | 15 052 |
| Q2.2 | 15 612 | 14 571 | 16 855 | 15 302 |
| Q2.3 | 3 004 | 1 372 | 1 161 | 1 222 |
| Q2.4 | 53 878 | 54 428 | 60 027 | 66 466 |
| Q2.5 | 12 317 | 10 866 | 15 152 | 12 307 |
| Q2.6 | 267 085 | 314 261 | 364 834 | 276 618 |
| Q2.7 | 316 104 | 174 942 | 258 562 | 224 593 |
| Q2.8 | 56 441 | 32 066 | 22 653 | 136 346 |
| Q2.9 | 26 258 | 27 780 | 30 884 | 129 987 |
| Q2.10 | 27 419 | 22 312 | 33 578 | 29 252 |
| Q2.11 | 1 072 | 1 012 | 152 810 | 148 624 |
| Q2.12 | 55 770 | 90 259 | 62 950 | 99 112 |
| Q2.13 | 61 348 | 53 457 | 69 540 | 62 009 |
| Q2.14 | 241 528 | 165 588 | 218 144 | 239 615 |
| Q2.15 | 403 500 | 345 357 | 481 573 | 485 358 |
| Q2.16 | 527 478 | 448 956 | 1 882 | 503 093 |
| Q2.17 | 445 902 | 499 908 | 608 445 | 459 201 |
| Q2.18 | 44 433 | 31 976 | 26 659 | 33 848 |
| Q2.19 | 56 091 | 55 410 | 52 876 | 57 072 |
| Q2.20 | 62 800 | 69 260 | 56 280 | 71 623 |
| **Gain** | **0%** | **9.8%** | **5.4%** | **-13.9%** |

Table 8: DW2 results

Table 9 finally presents the execution time (in milliseconds) of each of the queries Q3.1 to Q3.20 on data warehouse DW3, using each of the index configurations IC0, IC1, and IC3 (IC2 is not applicable on a star schema). Table 9's last line also features the average gain in performance when using a given index configuration instead of no index. Table 9 confirms that star join indices (configuration IC3) are the best choice on a star schema, as they are designed to be. This is especially visible with queries that return large results such as Q3.2, Q3.15 or Q3.18.

| Query | IC0 | IC1 | IC3 |
|---|---|---|---|
| Q3.1 | 2 603 | 1 922 | 731 |
| Q3.2 | 497 125 | 370 353 | 279 882 |
| Q3.3 | 12 228 | 2 183 | 1 923 |
| Q3.4 | 15 031 | 2 874 | 3 605 |
| Q3.5 | 14 411 | 3 185 | 2 704 |
| Q3.6 | 10 265 | 4 316 | 3 706 |
| Q3.7 | 6 529 | 4 266 | 6 499 |
| Q3.8 | 12 128 | 3 555 | 3 064 |
| Q3.9 | 16 984 | 14 020 | 17 455 |
| Q3.10 | 5 107 | 2 905 | 4 156 |
| Q3.11 | 6 730 | 4 076 | 6 740 |
| Q3.12 | 17 806 | 14 460 | 9 073 |
| Q3.13 | 7 400 | 9 184 | 7 080 |
| Q3.14 | 3 185 | 3 555 | 3 195 |
| Q3.15 | 173 960 | 92 983 | 83 800 |
| Q3.16 | 5 478 | 2 714 | 2 654 |
| Q3.17 | 53 076 | 33 839 | 35 441 |
| Q3.18 | 576 649 | 733 004 | 529 472 |
| Q3.19 | 802 | 811 | 610 |
| Q3.20 | 2 353 | 2 063 | 2 344 |
| **Gain** | **0%** | **9.3%** | **30.3%** |

Table 9: DW3 results

As a conclusion, we showed with these experiments how DWEB could be used to evaluate the performances of a given DBMS when executing decision support queries on several data warehouses. We underlined the critical nature of index choices and how they should be guided by both the data warehouse architecture and contents. However, note that, from a sheer performance point of view, these experiments are not wholly significant. For practical reasons, we indeed generated relatively small data warehouses and did not conduct real full-scale tests. Furthermore, our experiments do not do justice to Oracle, since we did not seek to achieve the best performance. For instance, we did not combine different types of indices. We did not use any knowledge about how Oracle exploits these indices either. Our experiments are truly sample usages for DWEB.

## 8. Conclusion and perspectives

We aimed in this paper at helping data warehouse designers to choose between alternate warehouse architectures and performance optimization techniques. For this sake, we proposed a performance evaluation tool, namely a benchmark (or benchmark generator, as it may be viewed) called DWEB (the *Data Warehouse Engineering Benchmark*), which allows users to compare these alternatives.

To the best of our knowledge, DWEB is currently the only operational data warehouse benchmark. Its main feature is that it can generate various ad-hoc synthetic data warehouses and their associated workloads. Popular data warehouse schemas, such as star schemas, snowflake schemas, and constellation schemas can indeed be achieved. We mainly view DWEB as an engineering benchmark designed for data warehouse and system designers, but it can also be used for sheer performance comparisons. It is indeed possible to save a given warehouse and its associated workload to run tests on different systems and/or with various optimization techniques.

This work opens up many perspectives for developing and enhancing DWEB toward Gray's relevance objective. First, the warehouse metamodel and query model were deliberately simple in this first version. They could definitely be extended to be more representative of real data warehouses. For example, the warehouse metamodel could feature many to many relationships between dimensions and fact tables, and hierarchy levels that are shared by several dimensions. Our query model could also be extended with more complex queries such as nested queries that are common in OLAP usages. Furthermore, it will be important to fully include the ETL process into our workload, and the specifications of TPC-DS and some other existing studies (Labrinidis & Roussopoulos, 1998) should help us.

We have also proposed a set of parameters for DWEB that suit both the models we developed and our expected usage of the benchmark. However, a formal validation would help selecting the soundest parameters. More experiments should also help us to evaluate the pertinence of our parameters and maybe propose sounder default values. Other parameters could also be considered, such as the domain cardinality of hierarchy attributes or the selectivity factors of restriction predicates in queries. This kind of information may indeed help designers to choose an architecture that supports some optimization techniques adequately.

We assumed in this paper that an execution protocol and performance metrics were easy to define for DWEB (e.g., using TPC-DS' as a base) and focused on the benchmark's database and workload model. However, a more elaborate execution protocol must definitely be designed. In our experiments, we also only used response time as a performance metric. Other metrics must be envisaged, such as the metrics designed to measure the quality of data warehouse conceptual models (Serrano *et al.*, 2003; Serrano *et al.*, 2004). Formally validating these metrics would also improve DWEB's usefulness.

Finally, we are also currently working on warehousing complex, non-standard data, such as multimedia, multistructure, multisource, multimodal, and/or multiversion data (Darmont *et al.*, 2005b). Such data may be stored as XML documents. Thus, we also plan a "complex data" extension of DWEB that would take into account the advances in XML warehousing (Nassis *et al.*, 2005; Rusu *et al.*, 2005).

## Acknowledgements

The authors would like to thank Sylvain Ducreux, Sofiane Guesmia, Bruno Joubert, Benjamin Mouton and Pierre-Marie Penin for their important contribution to DWEB's implementation and testing; as well as this paper's reviewers, for their invaluable comments.